\documentclass{ws-ijmpa}

\begin{document}

\markboth{F. Cianfrani, G. Montani}
{Synchronous Quantum Gravity}

%
\catchline{}{}{}{}{}
%

\title{Synchronous Quantum Gravity
}

\author{FRANCESCO CIANFRANI}

\address{ICRA---International Center for Relativistic Astrophysics\\ 
Dipartimento di Fisica (G9),
Universit\`a  di Roma, ``Sapienza'',\\
Piazzale Aldo Moro 5, 00185 Rome, Italy.\\ 
francesco.cianfrani@icra.it}

\author{GIOVANNI MONTANI}

\address{ICRA---International Center for Relativistic Astrophysics\\ 
Dipartimento di Fisica (G9),
Universit\`a  di Roma, ``Sapienza'',\\ 
Piazzale Aldo Moro 5, 00185 Rome, Italy.\\ 
ENEA C.R. Frascati (Dipartimento F.P.N.),
Via Enrico Fermi 45, 00044 Frascati, Rome, Italy.\\
ICRANet C. C. Pescara, Piazzale della Repubblica, 10, 65100 Pescara, Italy.\\
montani@icra.it}

\maketitle

\begin{history}
\received{Day Month Year}
\revised{Day Month Year}
\end{history}

\begin{abstract}

The implications of restricting the covariance principle within a Gaussian gauge are developed both on a classical and a quantum level. Hence, we investigate the cosmological issues of the obtained Schr\"odinger Quantum Gravity with respect to the asymptotically early dynamics of a generic Universe. A dualism between time and the reference frame fixing is then inferred.


\end{abstract}

\ccode{PACS number: 04.60.-m}

\section{INTRODUCTION}

A canonical quantum theory for the gravitational field is performed starting from an Hamiltonian formulation, hence from the definition of a space-time splitting. A criticism to this approach is that it is based on the identification of a time-like vector field; but it is expected that, once the geometry has been quantized, only a probabilistic response can be given for a vector to be time-like or space-like. A physical way to introduce a space-time splitting of the space-time is to include in the dynamics a real fluid. By virtue of the matter-time dualism \cite{BK95,MM04}, the introduction of matter provides a solution to the time-problem in Quantum Gravity \cite{I92}.    
Here the link between the identification of a reference and the appearance of a matter term is investigated from the point of view of Lagrangian symmetries. In particular, we are going to discuss the dynamical implications of the covariance principle when it is restricted to a synchronous frame of reference. The Hamiltonian constraints will be inferred, showing how fixing the reference leads to the materialization of a dust-like contribution. Finally, the description of a generic cosmological model will be given, together with the picture of a quantum to classical transition for the adopted evolutionary approach.

\section{Dynamics of the gravitational field in a synchronous reference}

A synchronous (gaussian) reference frame is characterized by a metric tensor having the following fixed components $g_{00} = 1$ and $g_{0i} = 0$
($i=1,2,3$), as soon as a 3+1 splitting $y^\mu=y^\mu(t,x^i)$ of the full space-time manifold is performed. This feature implies $N = 1$ and $N^i = 0$ for the lapse function and for the shift vector, respectively, in the ADM formalism. 
It can be shown that the synchronous character is preserved by the following set of restricted coordinate transformations
\cite{KL63},

\begin{equation}
t^{\prime } = t + \xi(x^l)\,,
\quad 
x^{i^{\prime }} = x^i
+ \partial _j\xi \int h^{ij}dt + \phi ^i(x^l)
\, , \label{srp3bis}
\end{equation}

$\xi$ and $\phi^i$ being three generic space functions.

By restricting the invariance requirement for the Lagrangian, one finds the following set of constraints, replacing the super-Hamiltonian and the super-momentum ones,

\begin{equation}
H^* \equiv H - \mathcal{E}(x^l) = 0,\qquad H_i = 0,
\label{srp16}
\end{equation}

$\mathcal{E}$ being a scalar density of weight $1/2$, hence it can be written as $\mathcal{E}\equiv -2\sqrt{h}\rho (t,\; x^i)$,
 with $\rho$ a scalar function.
 
These expressions outline that, while the super-momentum constraint is not modified, in fact the invariance under 3-difeomorphisms still stands, the super-Hamiltonian is non-vanishing, in order to preserve the synchronous character of the reference. 


The modification induced on the form of constraints can be interpreted as the emergence of a dust fluid co-moving with the slicing, whose energy density is given by the function $\rho$. Einstein equations underlying such a picture read 
\begin{eqnarray}
\label{agp5}
G_{\mu \nu } = k \rho u_{\mu }u_{\nu }\\
\rho = -\frac{\mathcal{E}}{2\sqrt{h}}, 
\end{eqnarray}
$u_\mu$ being orthogonal to the space-like slicing.
 
One can think at this contribution as the physical realization of the synchronous reference. However, it is clear that we are not dealing with an external matter field since the energy density $\rho$ is not always positive and $\mathcal{E}(x^i)$ is fixed, once
initial conditions are assigned on a non-singular hypersurface.

\section{Synchronous Quantum Gravity}

We perform the quantization of the synchronous gravitational
field by promoting the 3-geometry
$h_{ij}$ and its conjugated momentum $\pi ^{ij}$ to operators acting on the state
function $\chi $ in a canonical way, {\it i.e.} 

\begin{eqnarray}
\label{qo}
h_{ij}\rightarrow \hat{h}_{ij}\\
\pi ^{ij}\rightarrow \hat{\pi }^{ij} = 
-\frac{i\hbar}{(2ck)^{3/2}}
\frac{\delta (\quad )}{\delta h_{ij}}.
\end{eqnarray}

The constraints are implemented according with the Dirac prescription, hence they read as follows

\begin{eqnarray}
\label{nhc}
{\hat{H}}^*\chi _{\mathcal{E}} = 0 
\, \quad \Rightarrow \, \quad 
\hat{H}\chi _{\mathcal{E}} =
\mathcal{E}\chi _{\mathcal{E}} \label{1}\\           
{\hat{H}}_i\chi _{\mathcal{E}} = 0 
\, .\label{2}
\end{eqnarray}

The condition (\ref{2}) implies that the wave functional 
depends on the 3-geometries $\{ h_{ij}\}$, not on 3-metrics. But a relevant achievement is the relation (\ref{1}): $H$ being the generators of time displacements, it fixes an evolutionary character for wave functional, which can be described by the Schr\"odinger equation 

\begin{equation}
i\hbar \partial _t\chi =
\int _{\Sigma ^3_t}\hat{H} d^3x  \chi 
\, .
\end{equation}

Therefore, the quantum features of the dust contribution outline its behavior as a clock-like matter. The next task is to find out a negative portion of the super-Hamiltonian spectrum, which allows to interpret the additional contribution as a physical matter field.  

In this respect, a change of variables is performed on the 3-metric and it is written as

\begin{equation}
h_{ij} \equiv \eta ^{4/3}u_{ij}
\, ,
\label{detvar}
\end{equation}

with $\eta \equiv h^{1/4}$ and $det u_{ij} = 1$.\\
This way the Hamiltonian density
takes the form

\begin{equation}
H = -\frac{3}{16}c^2 kp_{\eta}^2+
\frac{2c^2 k}{\eta ^2}u_{ik}u_{jl}p^{ij}p^{kl}
- \frac{1}{2k}\eta ^{2/3}V(u_{ij},\; \nabla \eta ,\; \nabla u_{ij})
\label{newh}
\, ,
\end{equation}
$p_{\eta }$ and $p^{ij}$ being the conjugate momenta to
$\eta $ and $u_{ij}$, respectively.

Here, the potential term $V$ comes from the 3-Ricci scalar and
$\nabla$ refers to first and second order spatial gradients.

In this scheme, equations (\ref{1}) can be written as

\begin{equation}
\label{newhaux}
\hat{H}\chi _{\mathcal{E}} =
\left\{ 
\frac{3}{128\hbar ck^2}\frac{\delta ^2}{\delta \eta ^2} -
\frac{1}{4\hbar ck^2\eta ^2}\Delta _{u}                          
- \frac{1}{2k}\eta ^{2/3}V(u_{ij},\; \nabla \eta ,\; \nabla u_{ij})
\right\}\chi _{\mathcal{E}}  = 
\mathcal{E}\chi _{\mathcal{E}}
\end{equation}
\begin{equation}
\Delta _{u} \equiv
\frac{\delta }{\delta u_{ij}}
u_{ik}u_{jl}
\frac{\delta }{\delta u_{kl}}
\, .
\end{equation}

This equation has a Klein-Gordon-like
structure, hence there exist solutions with negative values of
$\mathcal{E}$. 

It is worth noting that in a synchronous reference the metric determinant always
vanishes monotonically at a time
$t^*$ where all the geodesics lines cross each other, i.e
$\eta (t^*, x^i) = 0$, with $\partial _{t\rightarrow t^*}\eta > 0$ (as a consequence of the Landau-Raichoudhuri theorem).
This property supports
i)$\eta$ as an internal time, ii) to take the limit
$\eta  \rightarrow 0$, where the system 
(\ref{newhaux}) admits an asymptotic solution.
In fact, in this limit, the potential term turns out to be
drastically suppressed with respect to the
term $\Delta _{u}$, thus the dynamics of different spatial points decouples and the quantization can be performed in a local minisuperspace.\\
An approximated solution is provided in this scheme, point by point in space, by the following expression

\begin{equation}
\chi _{\mathcal{E}} =
\iota _{\mathcal{E}}(\eta , p)G_{p^2}(u_{ij})
\, ,
\label{iog}
\end{equation}

$\iota $ and $G_{p^2}$ satisfying the two equations
respectively

\begin{eqnarray}
\label{twoeig1}
\left\{ 
\frac{1}{\hbar ck^2}\frac{\delta ^2}{\delta \eta ^2} +
\frac{32p^2}{\hbar ck^2\eta ^2}                                    
\right\}\iota _{\mathcal{E}} = 
\mathcal{E}\iota _{\mathcal{E}}\\
\Delta _{u} G_{p^2} = -p^2G_{p^2}
\, .
\end{eqnarray}

The potential results to be negligible, also on a quantum level, when the condition below holds
\begin{equation}
\frac{p^2}{4\hbar ck^2\eta ^2}\gg \frac{1}{2k}\eta ^{2/3}\frac{1}{\Delta u}\int_{\Delta u} d^5u V.
\end{equation}
A wider set of wave packets are in agreement with this condition, approaching $\eta=0$. For instance, one of them is characterized by a deviation $\Delta u\sim1/\Delta p\gg 1$, where $\Delta p$ is a small uncertainty around the picked value $p$ ($p\gg\Delta p$). Indeed, this approach works only if $p$ is greater than a fiducial one $p_0\sim\Delta p$ and this is consistent with the prescription that a quantum-classical correspondence stands for high quantum numbers.  

By writing $\iota = \sqrt{\eta } \theta (\eta )$, the function
$\theta $ obeys the following equation for the negative part of the spectrum
$\mathcal{E} = -\mid \mathcal{E}\mid$

\begin{eqnarray}
\label{twoeig2}
\frac{\delta ^2\theta }{\delta \eta ^2} +
\frac{1}{\eta }\frac{\delta \theta }{\delta \eta } +
\left(\mid \mathcal{E}^{\prime }\mid  -
\frac{q^2}{\eta ^2}\right)
\theta = 0 \\                     
\mathcal{E}^{\prime } \equiv
\hbar ck^2\mathcal{E} \\
q^2 \equiv \frac{1}{4}\left(1 - 128p^2\right)
\, .
\end{eqnarray}

A solution is given for
$\mid p\mid < 1/(8\sqrt{2})$ by

\begin{equation}
\theta (\eta ,\; \mathcal{E},\; p) =
AJ_q(\sqrt{\mid \mathcal{E}^{\prime }\mid } \eta ) +
BJ_{-q}(\sqrt{\mid \mathcal{E}^{\prime }\mid }\eta )
\, ,
\label{bessel}
\end{equation}

$J_{\pm q}$ being the corresponding
Bessel functions, while $A$ and $B$ are two integration
constants.

Therefore, a quantum range of positive energies exists for the dust contribution near enough to the
``singular'' point $\eta = 0$. However, there remain some open points:
i) The existence of a stable ground level of negative
$\mathcal{E}$.
ii) The inclusion of spatial gradients
of the dynamical variables.
iii) The physical meaning of the limit
$\eta \rightarrow 0$.
iv) The behavior of $\mathcal{E}$ in the classical limit.


\section{Synchronous Quantum Cosmology}

Most of these issues can be solved in a generic inhomogeneous cosmological setting, where the 3-metric is given by 

\begin{equation}
h_{ij} = e^{q_a}\delta_{ad}
O^a_b O^d_c \partial _iy^b \partial _jy^c,\ \ \
a,b,c,d,\alpha,\beta=1,2,3,
\label{lag1x}
\end{equation}

with $q^a = q^a(x^l,t)$ and $y^b = y^b(x^l,t)$ six scalar
functions 
and $O^a_b = O^a_b(x^l)$ a $SO(3)$ matrix.

The dynamics of different points decouples near the singularity and the Schr\"odinger
functional equation splits to the sum of $\infty ^3$ independent point-like
contributions as follows \cite{KL63,BM04} (we denote by the subscript $x$ any minisuperspace quantity)

\begin{eqnarray}
\label{sch}
i\hbar \partial _t \psi _x = 
\hat{H}_x\psi _x =  \frac{c^2\hbar ^2k}{3} \left[ 
\partial _{\alpha }e^{-3\alpha }\partial _{\alpha}
-e^{-3\alpha }\left( \partial ^2_+ + \partial ^2_-  
\right) \right]\psi _x 
- \frac{3\hbar ^2}{8\pi } e^{-3\alpha }\partial ^2_{\varphi }\psi _x
- \nonumber\\ 
- \left( \frac{1}{2k\mid J\mid ^2}e^{\alpha } V(\beta _{\pm })
- \frac{\Lambda }{k}e^{3\alpha }\right)  \psi _x\\
\psi _x = \psi _x (t , \; \alpha , \; \beta _{\pm} , \; \varphi )
\, ,
\end{eqnarray}

where a cosmological constant $\Lambda$ and a scalar field $\varphi$ have been added to the dynamical description.

If an integral representation is taken for the
wave function $\psi _x$

\begin{eqnarray}
\label{exp}
\psi _x = \int 
d\mathcal{E}_x \mathcal{B}(\mathcal{E}_x)
\sigma _x(\alpha , \; \beta _{\pm } , \; \varphi ,\; \mathcal{E}_x)
exp \left\{ -\frac{i}{\hbar }\int _{t_0}^tN_x\mathcal{E}_xdt^{\prime }
\right\}\\
\sigma _x = \xi _x(\alpha , \; \mathcal{E}_x)
\pi _x(\alpha , \; \beta _{\pm } , \; \varphi )
\, , 
\end{eqnarray}
where $\mathcal{B}$ is 
fixed by the initial conditions at $t_0$, the dynamics is given by

\begin{eqnarray}
\label{eigenvp}
\hat{H}\sigma _x = \mathcal{E}_x \sigma _x\\ 
\left( -\partial ^2_+ - \partial ^2_-
- \frac{9\hbar ^2}{8\pi c^2k } \partial ^2_{\varphi }
\right) \pi _x
- \frac{3}{2c^2\hbar ^2k^2\mid J\mid ^2}e^{4\alpha } V(\beta _{\pm })\pi _x =
v^2(\alpha ) \pi _x\label{eigenvp2}\\ 
\left[ \frac{c^2\hbar ^2k}{3}\left( 
\partial _{\alpha }e^{-3\alpha }\partial _{\alpha } \xi _x
+ e^{-3\alpha }v^2(\alpha ) \right) +
\frac{\Lambda }{k}e^{3\alpha }\right] \xi _x
= \mathcal{E}_x\xi _x
\, . 
\end{eqnarray}

In the expression above, it has been neglected any
dependence of $\pi _x$ on $\alpha $ because, asymptotically to
the singularity ($\alpha \rightarrow -\infty$),
it has to  be of higher
order (adiabatic approximation).

Let us now consider wave packets which are flat over the
width $\Delta \beta \sim 1/\Delta v_{\beta }\gg 1$
($\Delta v_{\beta }$ being the standard deviation in the
momenta space). 

In the new variable
$\tau = e^{3\alpha }$,
the equation (\ref{eigenvp2}) reads 

\begin{equation}
\frac{c^2\hbar ^2k}{3} \left( 
9\frac{d ^2 }{d \tau ^2}
+\frac{v^2}{\tau ^2}  
\right) \xi _x + 
\frac{\Lambda }{k}\xi _x
= \frac{\mathcal{E}_ x}{\tau }\xi _x
\, .  
\label{ph9}
\end{equation} 

A solution to equation
(\ref{ph9}) is provided by

\begin{eqnarray}
\label{solint}
\xi _x = \tau ^{\delta }f_x(\tau ),\qquad
\delta = \frac{1}{2}\left( 1 \pm
\sqrt{1 - \frac{4}{9}v^2}\right)\hspace{5cm}\\
\label{ph10}
f = \mathcal{C}
e^{-\beta ^2\tau ^2 + \gamma \tau}, \qquad
\gamma = 2\mid \beta \mid
\sqrt{\delta + \frac{1}{2}                 
-\frac{1}{12L_{\Lambda }^2l_P^4\beta ^2}},\qquad
\frac{1}{L_{\mathcal{E}}l_P^2} =
6\delta \gamma 
\,, \end{eqnarray} 
$L_{\mathcal{E}} = \frac{\hbar c}{\mathcal{E}}$ being the
characteristic length associated to the Universe ``energy'', 
while $l_P \equiv \sqrt{\hbar ck}$
denotes the Planck scale length. However, the validity of the solution above requires the condition $\beta ^2\tau \ll \gamma =
2\sqrt{\delta + \frac{1}{2}                 
-\frac{1}{12L_{\Lambda }^2l_P^4\beta ^2}}
\mid \beta \mid$.\\

Hence, the 
quantum dynamics in a
fixed space point
({\it i.e.} over a causal portion of the Universe)
is described, in the considered approximation
($\tau \ll 1$), by a free wave-packet for the variables
$\beta _{\pm }$ and $\varphi$  and by a profile in $\tau $
which has a maximum in
$\tau = (\gamma + \sqrt{\gamma ^2 + 8\delta \beta^2})/4\beta^2$. 

If a lattice structure for the space-time is assumed on the Planckian
scale \cite{RS03}, we can take $\tau >l_{Pl}^3$ and, noting that $\delta\leq1$, we obtain

\begin{equation}
\mid \beta \mid \ll
\frac{\sqrt{6}}{l_P^3}
\sqrt{1 
-\frac{1}{18L_{\Lambda }^2l_P^4\beta ^2}}
\label{fininequa}
\, .
\end{equation}

Furthermore, to preserve the reality of $\mathcal{E}$, we have to impose that the square root above is real, hence we get
$\mid \beta \mid \ge \frac{1}{3\sqrt{2}L_{\Lambda }l_P^2}$.
Finally, from the request $L_{\Lambda }\gg l_P$, then
the inequality (\ref{fininequa}) can be written as

\begin{equation} 
\mid \mathcal{E}_ x\mid \ll \frac{c^2k \hbar ^2}
{l_{Pl}^3} \sim \mathcal{O}(M_{Pl}c^2)
\rightarrow L_{\mathcal{E}} \gg l_P
\, ,
\label{ph11}
\end{equation} 

$M_{Pl} \equiv \hbar /(l_{Pl}c)$ being the Planck mass.

Therefore, the existence of a cut-off implies that a ground state exists for the evolutionary approach. Hence it is a natural request to assume the Universe to approach this state during its evolution.

In order to infer the influence of the dust contribution on the cosmological evolution, we evaluate the associated critical parameter. Being the super-Hamiltonian a constant of motion and assuming the Universe to be in the ground state, we find $\rho _{\mathcal{E}} \ll \mathcal{O}((M_{Pl}c^2)/R_0^3)$, with $R_0 \sim \mathcal{O}(10^{28}cm)$ the present Universe radius of curvature. Since the actual critical density can be
expressed as
$\rho _c \sim \mathcal{O}(c^4/[GR_0^2(\Omega - 1)])$
(being $\Omega = 1 \pm \mathcal{O}(10^{-2})$ the total Universe
critical parameter), then we have

\begin{equation}
\Omega _{\mathcal{E}} \equiv
\frac{\rho _{\mathcal{E}}}{\rho _c} \ll
\mathcal{O}\left( \frac{10^{-2}GM_{Pl}}{c^2R_0} \right) \sim 
\mathcal{O}\left( \frac{10^{-2}l_{Pl}}{R_0} \right) \sim 
\mathcal{O}\left( 10^{-60}\right)
\, .
\label{crit}
\end{equation}

Therefore, the dust contribution cannot play the role of dark matter. This conclusion can be modified by the introduction of matter from the thermal bath, which could have a relevant interaction with the Universe in the trans-Planckian region (see the model addressed in
\cite{CM04,BM06}, where ultrarelativistic matter
and a perfect gas were included). 


\section{The quantum-classical transition}

We are going to discuss if a scenario can be defined where the transition to a classical regime happens, before the spatial curvature contribution becomes relevant. This way, the obtained spectrum of $\mathcal{E}$ would result to be a good approximation during the whole cosmological evolution.



We take as a semi-classical wave function the following one

\begin{equation}
\sigma = \mu (\tau )
exp\left\{ i\frac{\Sigma (\tau )}{\hbar }\right\}
\mathcal{P}(\tau, \; \beta _{\pm }, \; \varphi )
\label{scwf}
\, ,
\end{equation}
which means that a classical behavior is going to be inferred for the time-variable $\tau$ only.

Addressing the same approximations scheme fixed in
\cite{V89}, we obtain the semi-classical equations below in the limit $\hbar \rightarrow 0$, by substituting $\sigma$ into the eigenvalue
problem (\ref{eigenvp}),

\begin{eqnarray}
\label{ph131}
-3c^2k  
\left( \frac{ d\Sigma }{d\tau }\right) ^2 +
\frac{\Lambda }{k}
- \frac{\mathcal{E}}{\tau } = 0 \\
\label{ph132}\frac{d }{d \tau }
\left( \mu ^2 \frac{d \Sigma }{d \tau }
\right) = 0\\ 
i\hbar \partial _t\mathcal{P} = 
-\frac{1}{\tau (t)}\left[ \frac{\hbar ^2c^2k}{3}\left(\partial ^2_+
+ \partial ^2_- \right) 
+ \frac{3}{8\pi } \partial ^2_{\varphi }
\right] \mathcal{P} 
\label{redsch}
\,,\end{eqnarray}
 
where in the last equation it has been used the identification $p_{\tau } = d\Sigma /d\tau$, coming out from the first one (which is the Hamilton-Jacobi equation for the considered system).
It is worth noting that the evolution of the wave function $\mathcal{P}$ is described by a Schr\"odinger equation. 

A solution to equations (\ref{ph131}) and (\ref{ph132}) can be given in the region
$\tau \ll L_{\mathcal{E}}L_{\Lambda }^2$ by 

\begin{eqnarray}
\label{ph14}
\Sigma= 2\sqrt{\frac{\mid \mathcal{E}\mid}{3c^2k}\tau}\qquad 
\mu = \mathcal{D}\sqrt[4]{\tau }\qquad
\mathcal{D} = \mathcal{D}(x^i) 
\,,
\end{eqnarray}

and they are in agreement with the
approximation scheme. 

The dynamics for $\mathcal{P}$ is that of a 3-dimensional
non-relativistic free particle with the time variable
given by $\mathcal{T} = \int(dt/\tau(t))$.
Hence, if at the initial time $\mathcal{P}$ is in the form of a
sufficiently narrow
packet, then it will remain peaked over
the whole regime where our approximations hold. Therefore,
within the adopted evolutionary approach, \emph{it can be inferred a quasi-classical description for the Universe volume, together with a quantum-like Schr\"odinger dynamics for the anisotropies
and the scalar field}. This feature demonstrate that a quantum-classical transition can be defined, where the dust contribution although being relevant during the Planckian regime, nevertheless it becomes negligible in the classical phase.

An open point remains the transition to a classical description for anisotropies and the scalar field. In this respect it could be significant the scalar curvature, which here has been neglected by the approximation scheme. In fact, such potential terms must approach a quadratic dependence on these variables, which allows to construct stable coherent states.


\end{document}